\documentclass[english,aps,twocolumn]{revtex4}
\usepackage[T1]{fontenc}
\usepackage[latin1]{inputenc}
\usepackage{graphicx}

\makeatletter

\usepackage{babel}
\makeatother
\begin{document}

\preprint{This line only printed with preprint option}

\title{Deformation properties of the BCP energy density functional}

\author{L.M. Robledo}

\email{luis.robledo@uam.es}

\affiliation{Dep. F\'\i sica Te\'orica C-XI, Universidad Aut\'onoma de Madrid, 28049 Madrid,
Spain}

\author{M. Baldo}

\affiliation{Instituto Nazionale di Fisica Nucleare, Sezione di Catania, Via Santa
Sofia 64, I-95123 Catania, Italy}

\author{P. Schuck}

\affiliation{Institut de Physique Nucl\`eaire, IN2P3-CNRS, UMR8608, F-91406 Orsay, France}
\affiliation{Universit\'e Paris-Sud, F-91406 Orsay, France}

\author{X. Vi\~nas}

\affiliation{Departament d'Estructura i Constituents de la Mat\`eria and Institut de Ci\`encies del Cosmos, Universitat
de Barcelona, Diagonal \emph{647}, 08028 Barcelona, Spain}

\begin{abstract}
We explore the deformation properties of the newly postulated BCP
energy density functional (EDF). The results obtained for three 
isotope chains of Mg, Dy and Ra are compared to the available experimental data as well
as to the results of the Gogny-D1S force. Results for the fission
barrier of $^{240}$Pu are also discussed.
\end{abstract}


\maketitle

\section{Introduction}

In a recent work \cite{BCP} we have shown that a fully microscopic input from 
nuclear and neutron Equation of State (EOS) calculations \cite{BMSV} 
complemented by additional terms accounting for finite size effects, to 
nuclear Density Functional Theory (DFT) can be very successfull. The density 
functional for the ground state energy $E$ of a nucleus, we considered in 
\cite{BCP}, is of the form
\begin{equation}
E = T_0 + E^{s.o.} + E_{int}^{\infty} + E_{int}^{FR} + E_C
\end{equation}
where $T_0, E^{s.o.}, E_C$ are the standard expressions for kinetic 
energy, spin-orbit term, and Coulomb energy. The dependence on proton 
and neutron densities in 
$E_{int}^{\infty}$ is the one obtained from the microscopic calculation 
\cite{BMSV} and treated in Local Density Approximation (LDA). We added a 
Finite Range (FR) term to account for a proper description of the 
nuclear surface:
\begin{eqnarray}
E_{int}^{FR}[\rho_n,\rho_p ] &=&
\frac{1}{2}\sum_{t,t'}\int
\int d^3r d^3r'\rho_{t}({\bf r})
v_{t,t'}({\bf r}-{\bf r'})\rho_{t'}({\bf r'} )
\nonumber \\
&-& \frac{1}{2}\sum_{t,t'}
\gamma_{t,t'}\int
d^3r {\rho_{t}({\bf r})} \rho_{t'}({\bf r})
\label{eq7} 
\end{eqnarray}
with $t$=proton/neutron and $\gamma_{t,t'}$ the volume integral of 
$v_{t,t'}(r)$. The second term in the r.h.s. of Eq (\ref{eq7}) is introduced to preserve
the nuclear matter properties of the microscopic calculation in the bulk. 
For the finite range form factor $v_{t,t'}(r)$ we made a 
simple Gaussian ansatz, that is $v_{t,t'} = V_{t,t'}e^{-r^2/r_0^2}$. We 
chose a minimum of three open parameters: $V_{p,p}=V_{n,n}=V_L$, 
$V_{n,p}=V_{p,n}=V_U$, and $r_0$. The values of these parameters together 
with 
the spin-orbit strength can be found in \cite{BCP}. They were adjusted to
reproduce the ground-state energy of some 
spherical nuclei. For the pairing part of the interaction we simply take
the density dependent delta force studied in \cite{Ga.99} for effective mass equal
to the bare one.
With this so-called Barcelona-Catania-Paris (BCP) 
functional excellent results for 161 even-even spherical nuclei with rms 
values for ground state energies and charge radii, 
comparable with the most performant 
functionals on the market \cite{D1S,Berger.89,NL3,sly4}, were obtained.

In this paper we continue investigating the properties of the BCP functional 
and explore how  deformation properties of nuclei are described. We will find that the 
performance is again excellent, as well in comparison with experiment as 
in comparison with the results of the very successfull Gogny D1S force 
\cite{D1S}. We have computed the Potential Energy Surfaces (PES)
as a function of the axially symmetric quadrupole deformation of several
isotopes of the Magnesium, Dysprosium and Radium species with the idea in mind 
of covering different regions and different nuclear scenarios of the
Nuclide Chart. To finish these exploratory calculations we have considered
also the fission barrier of the heavy nucleus $^{240}$Pu which is very sensitive
to quadrupole deformation properties.

\section{Methods}

As it is customary in this kind of calculations the HFB equation has
been recast as a minimization process of the energy density functional
where the HFB wave function of the Bogoliubov transformation
\cite{RS80} is chosen  to minimize the energy.
The variational set of HFB wave functions is given by means
of the standard Thouless parametrization \cite{RS80}. The minimization process 
is performed by using the gradient method as it allows an easy and efficient 
implementation of constraints by using the technique
of Lagrange multipliers. For the present exploratory calculations
we have restricted the calculation
to configurations preserving axial symmetry (but reflection symmetry
breaking is allowed) in order to reduce the computational effort. 
The quasiparticle operators are expanded in
a harmonic oscillator basis written as the tensor product of one oscillator
in the $z$ direction with oscillator frequency $\omega_{z}$ and
another in the perpendicular one and characterized by $\omega_{\perp}$.
Large enough basis (up to 26 shells in the $z$ direction in the case of fission)
and optimization of the harmonic oscillator lengths
have been used to guarantee good convergence for the values of all physical quantities.

In the calculations with the Gogny force the standard approximations
have been considered, namely the Coulomb exchange contribution to the energy
has been replaced by the Slater approximation and the Coulomb pairing
field has been neglected. The two body kinetic energy correction has
been fully taken into account. For details pertaining the implementation
of the HFB procedure and center of mass correction with the BCP functional 
see Ref. \cite{BCP}

\section{Results}

\begin{figure}
\includegraphics[width=0.9\columnwidth]{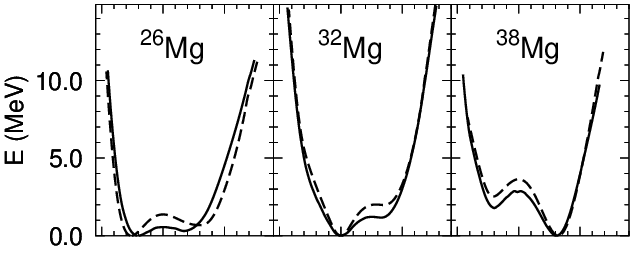}

\includegraphics[width=0.9\columnwidth]{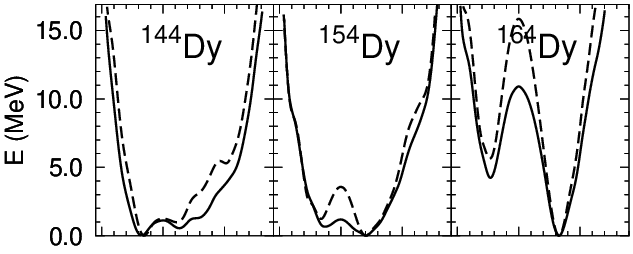}

\includegraphics[width=0.9\columnwidth]{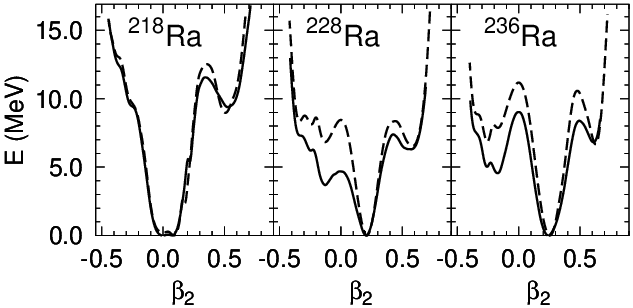}
\caption{The Potential Energy Surfaces (PES) of some representative nuclei as a function
of the $\beta_{2}$ deformation parameter (full line, BCP1 functional; dashed one Gogny-D1S force). 
In all the cases the zero of the curves is referred to their  absolute minimum. \label{FIG:PES}}
\end{figure}

We have performed calculations of the potential energy surface (PES), as
a function of the mass quadrupole moment $Q_{20}$, of several isotopes of the nuclei
Magnesium, Dysprosium and Radium. The idea is to explore some representative nuclei 
distributed all over the  Nuclide Chart. We have compared the results of the calculations 
with the BCP energy density functional (parametrizations BCP1 and BCP2) with those obtained
by using the D1S Gogny force which is considered here as a benchmark. For
the Mg isotopes we have computed the even mass ones between $A=20$
and $A=40$ (covering the N=8 and N=20 neutron shell closures). 
For the Dysprosium isotopes we have computed the ones
from $A=140$ (N=74) up to $A=170$ (N=104). Finally, for the Radium isotopes we
have considered isotopes between $A=216$ (N=128) up to $A=236$ (N=148). In Fig. (\ref{FIG:PES})
we have plotted the PES (as a function of the $\beta_2$ deformation
parameter, defined as $\beta_2=\sqrt{\frac{4\pi}{5}}\frac{Q_{20}}{\langle r^2\rangle }$) 
of some representative nuclei. As the results obtained 
with BCP1 and BCP2 parametrizations are
almost identical, only the first ones are shown. The first noticeable fact is that 
in all the nuclei studied the two curves
look rather similar showing minima, maxima and 
saddle points almost at the same $\beta_2$ values.
However, the
relative energy of those configurations obtained with BCP1 is different 
from the one with the Gogny force. According to the number of neutrons we can distinguish
two different regions: the first one corresponds to neutron numbers 
greater than the mid-shell value, where there are differences in the
prolate side between Gogny and BCP which extend up to rather high $\beta_2$ values
of around 0.5. Examples of this behavior are the nuclei $^{26}$Mg (N=14), $^{32}$Mg (N=20)  
and $^{144}$Dy (N=78). The second region corresponds to neutron numbers lower than the
mid-shell value where the spherical configuration lies, in the Gogny case, at an energy  higher than
in the BCP results and the difference increases as N increases. This means that the
deformation energy $E_{\mathrm{def}}$, defined as the energy difference between the spherical
configuration and the deformed ground state, is larger for the Gogny force than for the BCP
functional. As a consequence of this behavior the excitation energy of the oblate 
minimum is higher with the Gogny interaction than with the BCP functional, 
reaching the difference roughly a factor of two, as in
the example shown of $^{228}$Ra. Examples of
this behavior shown in Fig. (\ref{FIG:PES}) are $^{154}$Dy, $^{164}$Dy, $^{228}$Ra and $^{236}$Ra.
Finally, for some nuclei close or at shell closure like $^{218}$Ra  shown in Fig. (\ref{FIG:PES})
the two PES are rather similar.
These systematic differences observed between Gogny and BCP1 are
probably a consequence of the different surface energy coefficient
$a_{S}$  (the values are \textbf{$a_{S}=17.74$ MeV}
and 17.84 MeV for BCP1 and BCP2 respectively \cite{Farine.07} and $a_{S}=18.2$ MeV
for Gogny D1S \cite{Berger.89}).
Another possible explanation for the differences in the excitation energy of prolate 
and oblate minima could be the lower level density obtained with the 
Gogny force ( effective mass ratio of 0.7) as compared with the one of the 
BCP functional (effective mass equal to the physical one) that makes 
shell gaps stronger (see below). The deformation energy $E_\mathrm{def}$ also depends on the
amount of pairing correlations (see, for example Ref. \cite{Rutz.99}) in such a way that
the stronger the pairing correlations are the smaller the value of $E_\mathrm{def}$. It turns out that  the particle-particle correlation energy
$E_{pp}=\mathrm{Tr} (\Delta\kappa)$, which is a measure of pairing correlations, is typically
30 to 40 \% stronger in the Gogny-D1S case. However, its effect on  $E_\mathrm{def}$ is not strong enough as to overcome
the other effects mentioned above that lead to an increase of the deformation energy of Gogny D1S with respect to BCP.

Concerning the physics, we observe in Fig. (\ref{FIG:PES}) how shape coexistence in $^{26}$Mg with
its oblate ground state appears in both the BCP and D1S calculations.
We also observe the shoulder precursor of the deformed ground state
after angular momentum projection in $^{32}$Mg as well as the prolate
ground state that develops in heavier than $^{32}$Mg isotopes as
a consequence of neutrons populating the $fp$ shell (see $^{38}$Mg). In the case
of Dy we have two examples of shape coexistence with an oblate ground
state ($^{144}$Dy) and a prolate one ($^{154}$Dy) as well as a well
deformed system like $^{164}$Dy. Shape coexistence also appears in $^{142}$Dy
(not shown) but in this case BCP predicts a prolate g.s. whereas Gogny predicts an
oblate one. Finally in the lower row
of Fig. (\ref{FIG:PES}) we have several Ra isotopes ranging from
the spherical $^{218}$Ra showing an excited superdeformed minimum, to the
reasonably well deformed $^{228}$Ra and $^{236}$Ra where 
the excitation energies of the oblate minima are quite a bit higher in the 
case of the Gogny force than with the BCP functional.

\begin{figure}
\includegraphics[width=0.95\columnwidth]{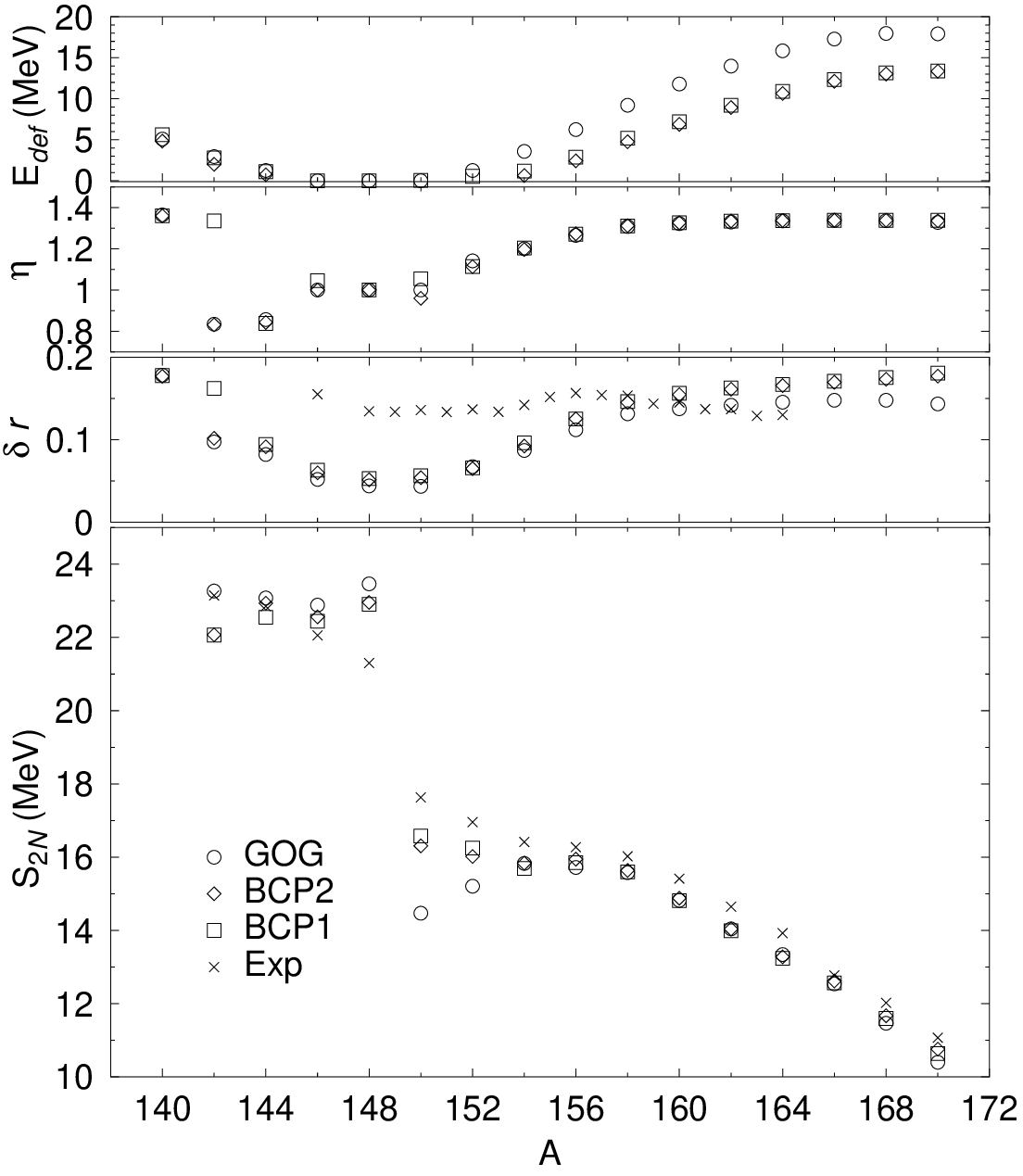}

\caption{Mean field results for the selfconsistent minimum of all the Dy isotopes
considered. In the lower panel the two neutron separation energy $S_{2\mathrm{N}}$ is
shown against the mass number of the isotopes along with the experimental
data. In the next panel the quantity $\delta r=\langle r^{2}\rangle^{1/2}-\sqrt{3/5}\times1.2\times A^{1/3}$
is depicted. In the following panel the axis ratio $\eta$ (see text for
details) is represented. Finally, in the upper panel the 
deformation energy $E_\mathrm{def}$ is plotted.\label{FIG:DyRES}}
\end{figure}

\begin{figure}
\includegraphics[width=0.9\columnwidth]{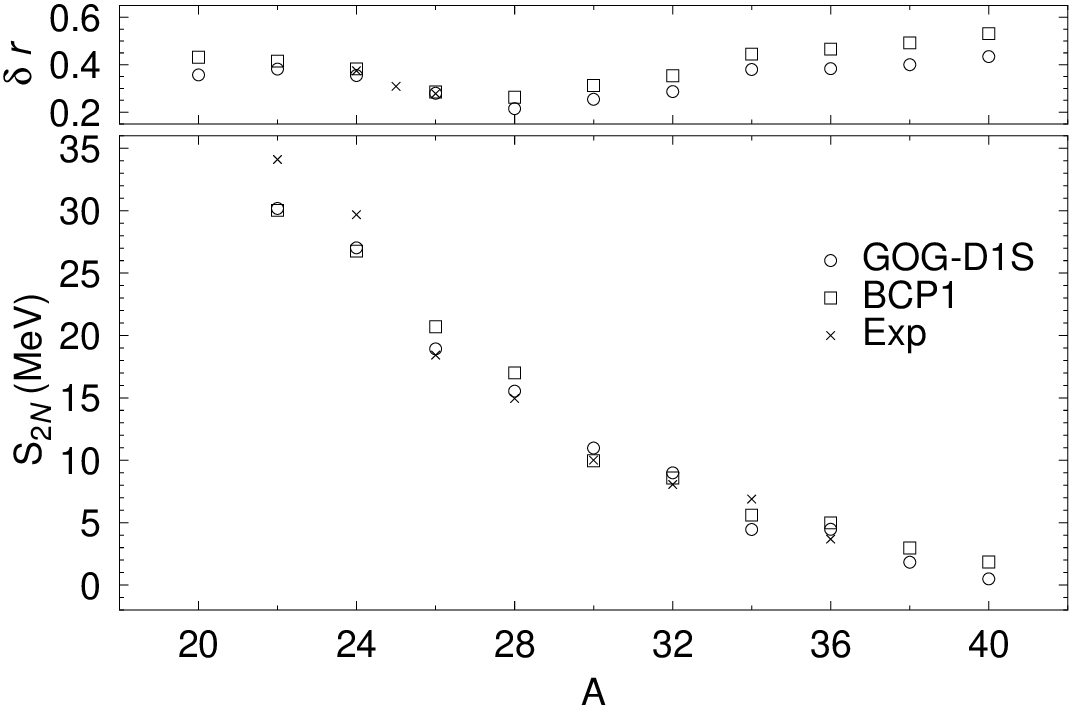}
\caption{Same as Fig. (\ref{FIG:DyRES}) but for the Mg isotopes and only
the two lower panels. \label{FIG:MgRES}}
\end{figure}
In Fig (\ref{FIG:DyRES}) we have displayed the magnitude of several physical
quantities corresponding to the ground state of the computed Dy isotopes, using the
two BCP functionals and the D1S force. In the lower panel of this figure
the two neutron separation energies are plotted. The typical discontinuity
at the semi-magic nucleus $^{148}$Dy (N=82,Z=66) is clearly observed.
As usual, our  predictions do not compare well with experiment
around that point because of the impact of missing
correlations in the binding energies of transitional (not well deformed) systems. In well
deformed systems the agreement with experiment is much better for all three cases considered here.  
The absolute values of the binding energies also agree well in all the cases and,
as an example, we can mention that in a calculation with a basis of
15 shells the binding energies of $^{160}$Dy are -1305.894 MeV for BCP1,
-1305.607 MeV for BCP2, -1304.288 MeV for D1S whereas the experimental value is -1309.457 MeV. 
The theoretical results do not include any
kind of correlation energy beyond mean-field like the rotational energy correction that
can be estimated to be 2.18 MeV for BCP and 3.16 MeV for D1S. In
the next panel of Fig (\ref{FIG:DyRES}) a quantity related to the
mean square radius, namely $\delta r=\langle r^{2}\rangle^{1/2}-\sqrt{3/5}\times1.2\times A^{1/3}$,
is plotted. We observe that the three theoretical predictions compare
rather well with experimental data and surprisingly the radii are closer to experiment
in well deformed nuclei which were not considered in the original
fit \cite{BCP}. The different result obtained in $^{142}$Dy for BCP1 and BCP2 (and D1S) 
is due to the almost degenerate
oblate and prolate minima in this nucleus: with BCP1 the ground state
is prolate whereas it is oblate with BCP2 and Gogny-D1S. In the next
panel the axis ratio $\eta=\left(\langle z^{2}\rangle/\langle x^{2}\rangle\right)^{1/2}$
which is a measure of deformation ($\beta_2=\sqrt{\frac{4\pi}{5}}\frac{\eta^2-1}{\eta^2+2}$)
($\eta > 1$ for prolate deformation, $\eta < 1$ for oblate
deformations and $\eta=1$ for spherical states) is plotted. As
in previous cases, the agreement between the three theoretical results
is very good confirming what was said in discussing  Figure (\ref{FIG:PES})
about the coincidence of the position of maxima and minima of the PES.
Finally in the upper panel the deformation energy $E_\mathrm{def}$  is shown. 
The Gogny-D1S values for this magnitude are systematically larger by a few MeV than
the BCP ones as discussed previously.

\begin{figure}
\includegraphics[width=0.9\columnwidth]{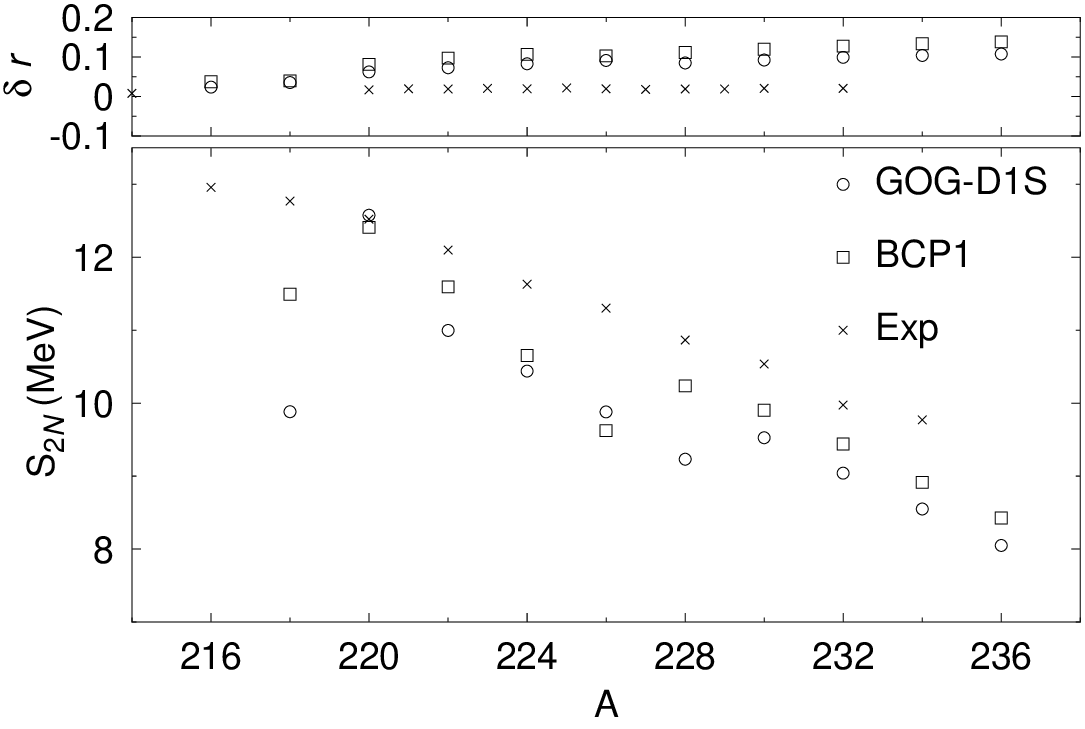}
\caption{Same as Fig. (\ref{FIG:MgRES}) but for the Ra isotopes. \label{FIG:Ra}}
\end{figure}

The results for the selfconsistent ground state in the Mg isotopes
is depicted in Fig. (\ref{FIG:MgRES}) where we have only plotted
the curves for $S_{2N}$ and $\delta r$.
The two additional quantities shown in the case of the Dy isotopes have a similar
behavior for the Mg nuclei and are not presented here. Again, we 
observe for the $S_{2N}$ a reasonably good agreement with experiment 
which is of the same quality for the three schemes. The theoretical
radii also look rather similar in all the theoretical  approaches and compare well
with the scarce experimental values.

In Fig. (\ref{FIG:Ra}) we show the selfconsistent results for 
the $S_{2N}$ of the
Ra isotopes. The results for the isotopes $^{220-226}$Ra include
octupole deformation in their ground state with $\beta_{3}$ values
of the order of 0.15 for all the cases considered. It is not the
scope of the present work to discuss octupole deformation in detail
but it is worth pointing out that also this deformation multipole
comes out quite the same independent of the force. The theoretical $S_{2N}$
results look rather similar  except for the nucleus
$^{218}$Ra which is close to the semi-magic $^{214}$Ra.
As it was discussed before, the two neutron separation energies
have a somewhat erratic behavior around semi-magic configurations. Most surprising
is the region around A=226-228 where the theoretical predictions move
away from experimental values. Concerning
the radii a good agreement between the results obtained with BCP, Gogny and the experiment is
observed.

\begin{figure}
\includegraphics[width=0.95\columnwidth]{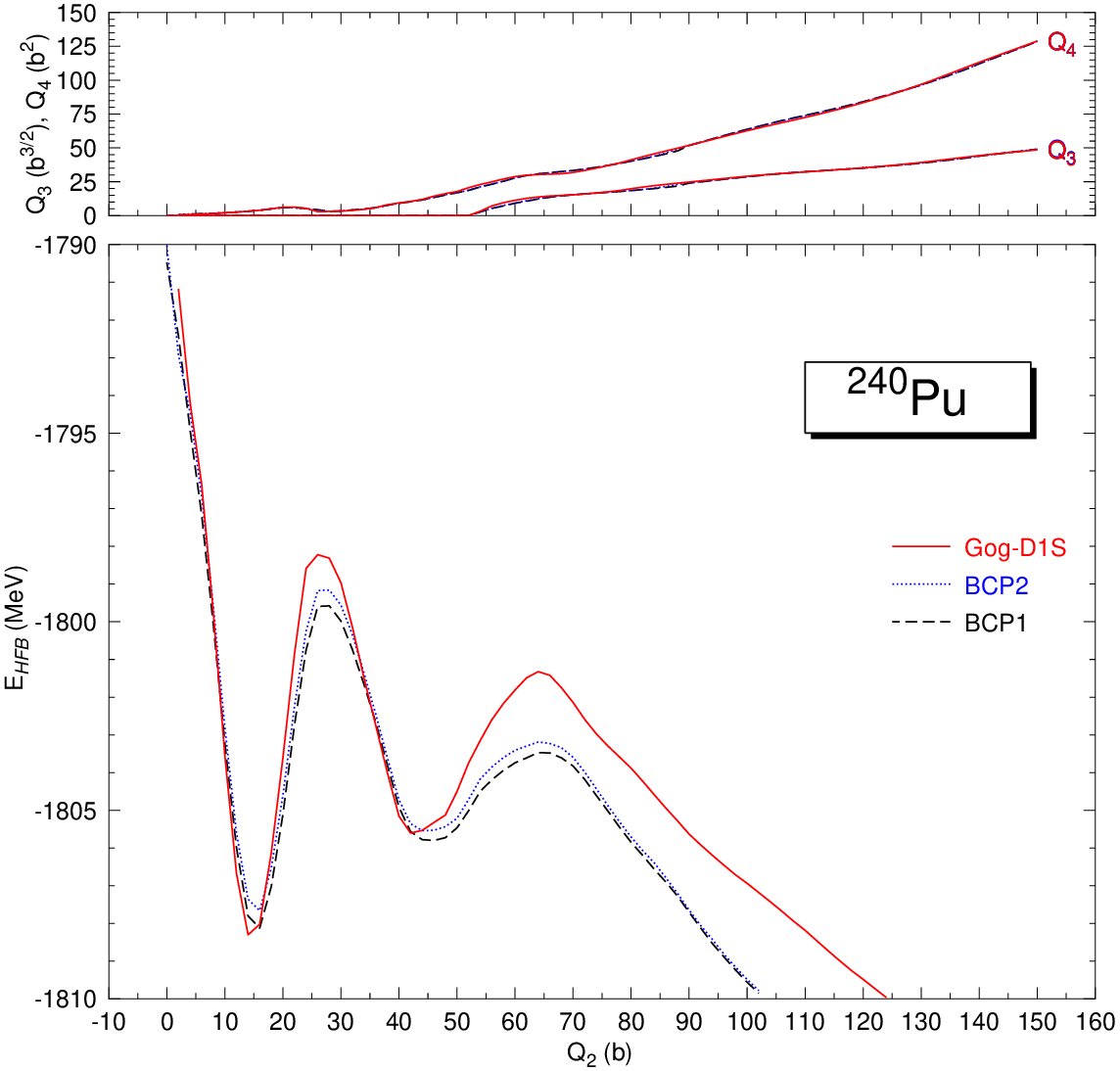}
\caption{Fission properties for the nucleus $^{240}$Pu computed with the
two parametrizations of the BCP functional as well as with the Gogny-D1S
force. In the lower panel the potential energy surfaces are depicted
as a function of the quadrupole moment. 
In the upper panel the octupole and hexadecapole
moments are presented. \label{FIG:FISSION}}
\end{figure}

In Figure (\ref{FIG:FISSION}) we display the potential energy surface (PES) 
as a function of the quadrupole moment corresponding to the fission process of the
nucleus $^{240}$Pu computed using the two BCP functionals and the Gogny force. As it is
customary in this kind of calculations \cite{Warda.02} the rotational energy correction
computed with the Yoccoz moment of inertia has been included in the PES. 
It can be concluded from this figure (lower panel) that the PES
obtained with the BCP1 and BCP2 functionals are very similar and both closely follow the shape of the 
Gogny D1S PES up to the second minimum. The heights of the fission barriers provided
by both the BCP1 and BCP2 functionals are lower
than the one calculated with D1S, as it is expected from the lower surface
energy values of BCP reported above. The smaller values of the fission barriers for
the BCP functional results go in
the right direction as compared to the experimental estimations \cite{Bjo.80,Firestone.96} but
the effect of  triaxiality in the first barrier in the case of the BCP functional 
remains to be studied.
The impact in the spontaneous fission half life $t_{\mathrm{SF}}$ of $^{240}$Pu is also uncertain as
this quantity not only depends on the topology of the PES but also on the
collective inertia that has not been considered here. Preliminary calculations not including triaxiallity
(therefore producing too high values for the half life) yield $t_{\mathrm{SF}}=1.2\times 10^{28}$s for
BCP1, $1.1\times 10^{27}$s for BCP2 and $1.5\times 10^{26}$s for D1S. Therefore it seems that
the larger values of the collective inertia obtained for the BCP functionals (consistent with their
lower pairing correlations) somehow counteract their lower fission barriers
providing a longer half life. 
Although the shapes involved
in fission are mainly characterized by its quadrupole moment, higher
multipole deformations can be important for describing some fine details of the nuclear dynamics.
In the top panel of Figure (\ref{FIG:FISSION})  the expectation
values of the octupole and hexadecapole moments are displayed as a function of
$Q_{20}$ and striking similarity between the 
three results is observed.

\begin{figure}
\includegraphics[width=0.45\columnwidth]{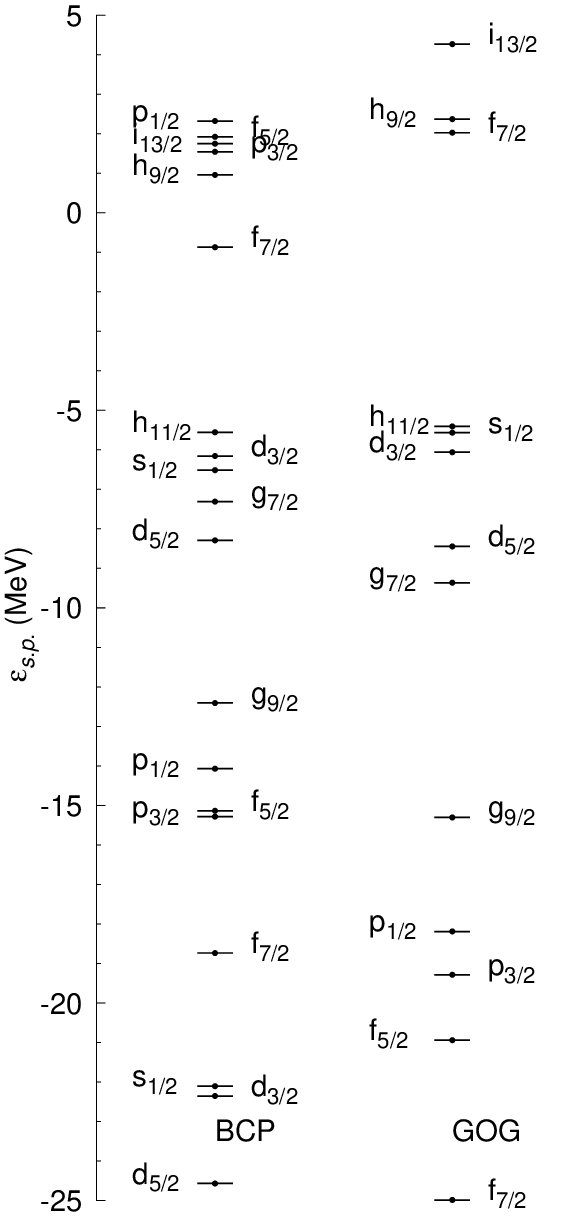}%
\includegraphics[width=0.45\columnwidth]{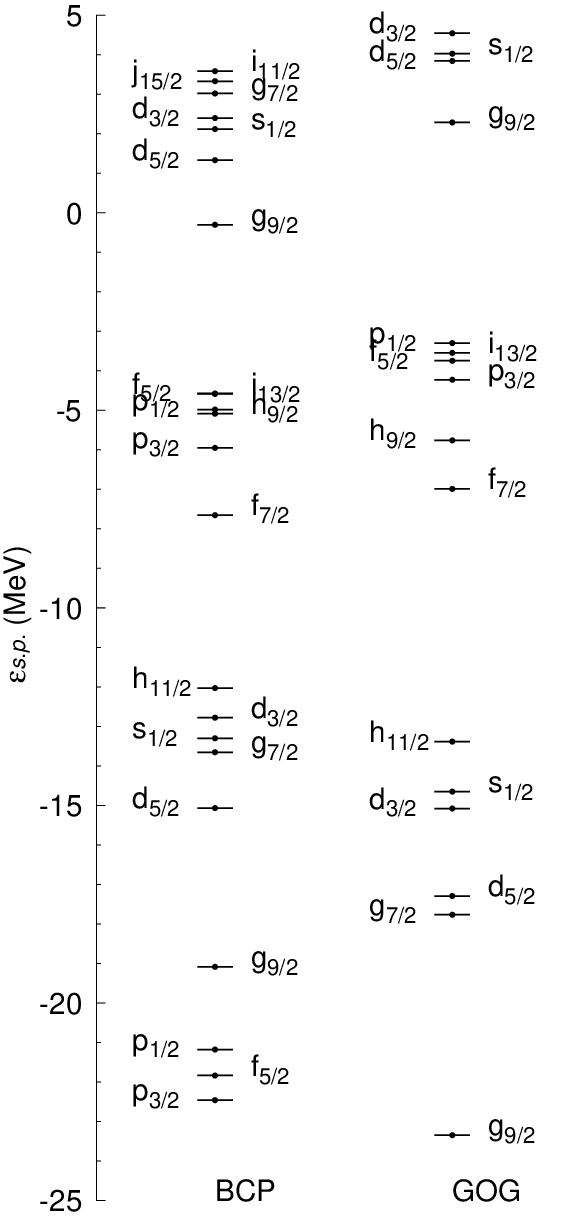}%
\caption{Spherical single particle energies for protons (left panel) and
neutrons (right panel) for the nucleus $^{160}$Dy. The results for the BCP1 functional
and the Gogny force (GOG) are given. To help identify the levels their labels 
alternate their position with respect
to the line assigned to them \label{FIG:SPE}}
\end{figure}

The similarity between the results obtained with the  BCP functionals and the Gogny 
force calls for a comparison of the underlying single particle structure that, as it 
is well known, is essential in determining the response of the system to deformation. 
The most thorough comparison would involve the analysis of the single particle energy 
plots as a function of quadrupole deformation (Nilsson diagrams) but this is a demanding task that is
deferred to a longer publication. We just want to mention here that the most relevant
feature of the Nilsson diagrams, namely the shell gaps determining the location of minima are
rather similar for both the BCP functionals and the Gogny force. This is so in spite of 
the different effective mass ratios $m^*/m$ in the two cases, one for the BCP functionals 
and 0.7 for Gogny-D1S, that imply a higher level density  in the former case. To illustrate
the different level densities we have plotted in Figure (\ref{FIG:SPE}) for the nucleus $^{160}$Dy the single
particle levels at sphericity ($Q_{20}=0$) for BCP1 and Gogny-D1S both for protons and 
neutrons. The higher level density of BCP's single particle levels is clearly observed. 
Results for BCP2 are not included, since they are very similar to the ones of BCP1.

\section{Conclusions}
We have analyzed by means of a few selected examples the performance
of the new BCP functional in what concerns quadrupole deformation.
The tests performed are quite demanding as they include fission barriers
for Pu or quadrupole properties over a wide range of isotopes in various
regions of the Nuclide Chart. We have compared the results of our
mean field calculations with experimental data whenever possible
and found good agreement comparable to the one obtained for spherical
magic or semimagic nuclei. Other quantities not directly related to
experiment like the topology of the potential energy surfaces have
been compared to the results of a well performant force and with long
tradition in the field, namely the Gogny-D1S force and the results
are comparable and of the same quality. Minor differences can
be attributed to a slightly lower surface tension for BCP compared to D1S 
and to the different pairing interactions used.
From these exploratory 
calculations we can conclude that the BCP functional with the
BCP1 and BCP2 parametrization can be used with confidence in
the study of nuclear properties related to deformation. Obviously
a more thorough study of the BCP functional with respect to deformation has to
be carried out as, e.g the study of octupole deformation, triaxiallity, collective inertia,
including the moment of inertia, etc. Work in this direction is underway.

\begin{acknowledgments}
This work has been partially supported by the CICyT-IN2P3 and CICyT-INFN collaborations.
L.M.R. acknowledges financial support from the DGI of the MEC (Spain)
under Project FIS2004-06697. X.V. acknowledges support from grants FIS2005-03142 of 
the MEC (Spain) and FEDER and 2005SGR-00343 of Generalitat de Catalunya.
\end{acknowledgments}


\begin{thebibliography}{1}
\bibitem{BCP}          M. Baldo, P. Schuck and X. Vi\~nas, 
                       nucl-th/0706.0658v2
\bibitem{BMSV}         M. Baldo, C. Maieron, P. Schuck and X. Vi\~nas,
                       Nucl. Phys. {\bf A736} (2004) 241 and references therein.
\bibitem{Ga.99}        E. Garrido, P. Sarriguren, E. Moya de Guerra and P. Schuck, 
                       Phys. Rev. {\bf C60}, 064312 (1999) 
\bibitem{D1S}          J.F. Berger, M. Girod and D.Gogny,
                       Comput. Phys. Commun. {\bf 63}, 365 (1991)
\bibitem{Berger.89}    J.F. Berger, M. Girod and D.Gogny,
                       Nucl. Phys. {\bf A502}, 85c (1989).
\bibitem{NL3}          G.A. Lalazissis, J. K\"oning and P.Ring,
                       Phys. Rev. {\bf C55}, 540 (1997).
\bibitem{sly4}         E. Chabanat, P. Bonche, P. Haensel, 
                       J. Mayer and R. Schaeffer,
                       Nucl. Phys. {\bf A627} (1997) 710; {\bf A635} (1998) 231
\bibitem{RS80}         P. Ring and P. Schuck, The nuclear many body problem (Springer, Berlin) (1980).
\bibitem{Farine.07}    M. Farine, private communication.
\bibitem{Rutz.99}      K. Rutz, M. Bender, P.-G. Reinhard and J.A. Maruhn, 
                       Phys. Lett. B468, 1 (1999).
\bibitem{Warda.02}     M. Warda, J.L. Egido, L.M. Robledo and K. Pomorski, 
                       Phys. Rev. C66, 014310 (2002)
\bibitem{Firestone.96} R.B. Firestone, V.S. Sirley, S.Y.F. Chu, C.M. Baglin, J. Zipkin, 
                       Table of isotopes, John Wiley and Sons, New York, 1996
\bibitem{Bjo.80}       S. Bjornholm, J.E. Lynn, 
                       Rev. Mod. Phys. 52 (1980) 725
\end{thebibliography}
\end{document}